\documentstyle[12pt]{article}
\def\psla{p{\raise1pt\hbox{$\!\!/$}}}
\def\dsla{\partial{\raise1pt\hbox{$\!\!\!/$}}}
\def\Dsla{D{\raise1pt\hbox{$\!\!\!/$}}}
\def\xsla{x{\raise1pt\hbox{$\!\!\!/$}}}

\def\ben{\begin{equation}}
\def\een{\end{equation}}
\def\bey{\begin{eqnarray}}
\def\eey{\end{eqnarray}}
\def\ba{\begin{array}}
\def\ea{\end{array}}

\def\xi0{\Xi^0}

\def\alphaxi0{\bar\alpha_\xi0}

\def\qqv{\langle{\bar q}q\rangle_{0}}
\def\uuv{\langle{\bar u}u\rangle_{0}}

\def\ddv{\langle{\bar d}d\rangle_{0}}

\def\ssv{\langle{\bar s}s\rangle_{0}}

\def\qq0v{\langle0\!\mid\!{\bar q}q\!\mid\! 0\rangle}
\def\qqf{\langle{\bar q}q\rangle_{F}}

\def\qc0f{\langle0\!\mid\!{\bar q}q\!\mid\!0\rangle_{F}}

\def\qsq0f{\langle0\!\mid\!{\bar q}\sigma_{\mu\nu}q\!\mid\!0\rangle_{F}}

\def\qgdq0f{\langle0\!\mid\!{\bar q}{\cal S}\gamma_{\mu}D_{\nu}q\!\mid\!0\rangle_{F}}

\def\qddq0f{\langle0\!\mid\!{\bar q}{\cal S}D_{\mu}D_{\nu}q\!\mid\!0\rangle_{F}}

\def\t{t}

\def\F{F_{\mu\nu}}

\def\nnbr{\nonumber}
\def\p0{p_0}
\def\mn{M_B}
\def\mb{M_B}

\def\gam3{\mbox{\boldmath{$\gamma$}}}
\def\e0{E_{0}(s_{0},s)}
\def\e1{E_{1}(s_{0},s)}
\def\e2{E_{2}(s_{0},s)}

\def\aplt{\kern0.3333em \raise 0.2ex \hbox{$<$}%
\kern-0.8em \lower0.8ex \hbox{$\sim$}%
\kern0.3333em}

\begin{document}


\baselineskip 30pt
\begin{center}
{\LARGE\bf Electric Polarizabilities of Neutral Baryons in the QCD Sum Rule}
\end{center}

\vspace{.5cm}

\begin{center}
{\large Tetsuo NISHIKAWA
\footnote{E-mail address nishi@nuc-th.phys.nagoya-u.ac.jp}
and Sakae SAITO
\footnote{E-mail address saito@nuc-th.phys.nagoya-u.ac.jp} 
}\\
{\it Department of Physics, Nagoya University, Nagoya 464-01, Japan}

\vspace{1.0cm}
{\large Yoshihiko KONDO\footnote{E-mail address kondo@kokugakuin.ac.jp}}\\
{\it Kokugakuin University, Tokyo 150, Japan}
\end{center}

\vspace{1.0cm}

\baselineskip 18pt

\noindent{\bf Abstract}\par
{\small We investigate the electric polarizabilities of neutral baryons 
using the method of QCD sum rules. 
The diagrams in the operator product
expansion are taken into account up to dimension 6.
We obtain different values of the polarizabilities among hyperons,
the ordering of which is given
$\bar\alpha_n>\bar\alpha_{\Xi^0}>\bar\alpha_{\Lambda}$.
The polarizability of ${\Sigma^0}$, $\bar\alpha_{\Sigma^0}$,
 has possibly a very small value.}

\newpage

Electric and magnetic polarizabilities, labeled ${\bar\alpha}$ and
${\bar\beta}$, are fundamental constants characterizing the hadron
structure. These constants measure the ease with which a hadron
acquires electric and magnetic dipole moments in external
electromagnetic fields [\ref{Lvov}]. While for the proton and neutron they have
long been studied through theoretically and experimentally
and recent measurements have established certain values 
[\ref{exp:Fed}-\ref{exp:Sch}],
very little is known about hyperon polarizabilities.
However with the hyperon beams at CERN and Fermilab,
the situation is expected to change. In particular the $\Sigma$
polarizabilities will be soon measured [\ref{Felmilab}].

This has induced a number of theoretical investigations.
In fact predictions have been made in the non-relativistic quark model 
[\ref{NRQM}],
the heavy baryon chiral perturbation theory [\ref{HBCHPT}] and the bound
state soliton model [\ref{BS-skyrme}]. 
In the non-relativistic quark model the results are
${\bar\alpha}_{\Sigma^+}=20.8$, ${\bar\alpha}_{\Sigma^-}=5.7$,
all in units of $10^{-4}{\rm fm}^{3}$.
The large difference in them can be traced back to the
different quark structure of the $\Sigma^+$ and $\Sigma^-$.
While in the $\Sigma^+$ the two quark flavors have opposite electric
charges, all valence quarks in the $\Sigma^-$ have the same charges
, which inhibit internal excitation. 
In agreement with the non-relativistic quark model,
the heavy baryon chiral perturbation theory (HBChPT) at leading order 
finds $\bar\alpha_{\Sigma^+}>\bar\alpha_{\Sigma^-}$
; namely, ${\bar\alpha}_{\Sigma^+}=9.4(8.8)$ and
${\bar\alpha}_{\Sigma^-}=6.3(5.9)$ in the same units for 
the axial-vector coupling constants $D=0.8(0.75)$
and $F=0.5(0.5)$.
Here the reason for this result is the kaon cloud contribution,
which is strongly suppressed for the $\Sigma^-$. 
In addition hyperon polarizabilities are smaller than the nucleon
polarizabilities 
($\bar\alpha_p=13.0(12.1)$, $\bar\alpha_n=11.4(10.5)$) in the HBChPT. 
The large value of $\bar\alpha_N$ in the HBChPT is due to the
large contribution of pion cloud. On the other hand 
the hyperon polarizabilities mainly come from the 
contribution of kaon cloud, whose mass is larger than that of pion
and inhibit internal excitation under the external field.

We use a completely alternative approach, the method of QCD sum rules 
[\ref{SVZ1},\ref{SVZ2}].
In a previous work [\ref{QSR}] 
we constructed the QCD sum rule for the electromagnetic
polarizabilities of the neutron 
and found that they are related with the vacuum
condensates of quark-gluon fields in a weak external constant electromagnetic
field. We have seen that the neutron electric polarizability is well described 
with the method. In this paper we extend the method to hyperons
and calculate the electric polarizabilities of neutral hyperons.

We consider a two-point correlation function in a weak external
constant electromagnetic field $F_{\mu\nu}$ as follows:
\begin{equation}
   \Pi_{F}(p)= i\int d^4x\,e^{ipx}\langle 0\!\!\mid \!\!
                       T[\eta_{B}(x){\bar\eta}_{B}(0)] \!\!
                                        \mid\!\! 0\rangle_{F},\label{twopoint}
\end{equation}
where $\eta_{B}(x)$ is an interpolating field constructed out of 
quark fields such that it has the quantum numbers of 
a baryon under consideration. 
Following Ioffe and Smilga [\ref{I&S}], we expand the correlation function
in powers of $F_{\mu\nu}$ up to order $F^2$ terms:
\begin{equation}
    \Pi_{F}(p)=\Pi^{(0)}(p)+\Pi^{(1)}_{\mu\nu}(p)F^{\mu\nu}
           +\Pi^{(2)}_{\mu\nu\rho\sigma}(p)F^{\mu\nu}F^{\rho\sigma}
           +{\cal O}(F^3). \label{tenkai}
\end{equation}
The leading term $\Pi^{(0)}(p)$ is the correlation function 
when the external field is absent.
$\Pi^{(1)}_{\mu\nu}(p)$ is known to be related with the magnetic 
moments of baryons. It is $\Pi^{(2)}_{\mu\nu\rho\sigma}(p)$ 
that is related with the electromagnetic
polarizabilities of baryons.
Requiring parity invariance, 
we can decompose $\Pi^{(2)}_{\mu\nu\rho\sigma}(p)$ into 
the following Lorentz structure:
\bey
\Pi^{(2)}_{\mu\nu\rho\sigma}(p)
&=&g_{\mu\rho}g_{\nu\sigma}\Pi_{S_{1}}(p^2)
  +p_{\mu}p_{\rho}g_{\nu\sigma}\Pi_{S_{2}}(p^2) \nnbr \\
& &+g_{\mu\rho}g_{\nu\sigma}\psla \Pi_{V_{1}}(p^2)
   +p_{\mu}\gamma_{\rho}g_{\nu\sigma}\Pi_{V_{2}}(p^2)
   +p_{\mu}p_{\rho}g_{\nu\sigma}\psla \Pi_{V_{3}}(p^2) \\
& &+\sigma_{\mu\lambda}p^{\lambda}p_{\rho}g_{\nu\sigma}\Pi_{T_{1}}(p^2)
   +i\gamma_{5}\epsilon_{\mu\nu\rho\lambda}
        [g^\lambda_{\;\sigma}\Pi_{P_{1}}(p^2)
         +p^{\lambda}p_{\sigma}\Pi_{P_{2}}(p^2)
        ]\cr
& &+i\gamma_{5}\epsilon_{\mu\nu\rho\lambda}
      [g^\lambda_{\;\sigma}\psla\Pi_{A_{1}}(p^2)
         +p^{\lambda}\gamma_{\sigma}\Pi_{A_{2}}(p^2)
         +\gamma^{\lambda}p_{\sigma}\Pi_{A_{3}}(p^2)
         +p^{\lambda}p_{\sigma}\psla\Pi_{A_{4}}(p^2)
      ].\nnbr
\eey
The invariant 
functions in the right hand side satisfy dispersion
relations:
\begin{equation}
    \Pi(p^{2})=\frac{1}{\pi}\int_{0}^{\infty} dt
                              \frac{{\rm Im}\Pi(t)}{t-p^{2}}
                             +{\rm subtraction\;\;terms},
\end{equation}
where the subscripts $S_{1}$ etc. are suppressed for brevity.
One can apply the Borel transform $\widehat{B}$ defined by  
\ben
\widehat{B}[\Pi(p^{2})] \equiv
     \lim_{\stackrel{\stackrel{-p^2\rightarrow\infty}{n\rightarrow\infty}}{\frac{-p^2}{n}=s}}
          \frac{(-p^2)^{n+1}}{n!}
          \left[ -\frac{d}{d(-p^{2})} \right]^n \Pi(p^{2}).
\een
Here $s$ denotes the squared Borel mass.
We make a Borel transform 
on both sides of the dispersion relations:
\ben
\widehat{B}[\Pi(p^{2})]
=\frac{1}{\pi}
         \int_{0}^{\infty}\! dt \,
            e^{-t/s} {\rm Im}\Pi(t).
        \label{BSR}
\een
Evaluating the left hand side
by an operator product expansion (OPE) in the deep Euclidean region
we obtain the Borel sum rules.

Let us next consider the physical content of the invariant functions
in the right hand side of Eq.(\ref{BSR}). 
The lowest intermediate state of 
the correlation function $\Pi(p)$ 
is expressed by the baryon propagator:
$ \Pi(p)=-\lambda_B^2/[\psla-\mn-\Sigma(p,F)]$,
where $\mn$ is the baryon mass, and $\lambda_B$ the coupling strength
of the interpolating field to the baryon. $\Sigma(p,F)$ is the self-energy part
under the influence of the background electromagnetic field,
and can be decomposed
into possible types of Lorentz structure.
The invariant functions of the self energy on the mass-shell ($p^2=\mb^2$) are
related to physical quantities of a baryon, such as the magnetic
moment, the polarizabilities, via an effective Lagrangian
introduced by L'vov [\ref{Lvov}].
This Lagrangian, which was constructed
by requiring Lorentz and gauge invariance, and
the discrete symmetries, describes the low-energy interaction of 
spin-$1/2$ composite particles with a soft electromagnetic field.
The amplitude of low energy Compton scattering 
is reproduced from the Lagrangian.
For the details see our preceding paper Ref.[\ref{QSR}].
We find that the invariant functions $\Pi_{S_i}$ and $\Pi_{V_i}$
are related with the polarizabilities in the following way:
\bey
\Pi_{S_1}(\t)
&=&-\lambda_{B}^2\left\{
   {1\over(\t-\mn^2)^3}[4\mn^3\Sigma_{1}(\t)]
   +{1\over(\t-\mn^2)^2}[3\mn\Sigma_{1}(\t)+2\mn^2\Sigma_{S}(\t)]\right.\cr
& &\qquad\left.+{1\over\t-\mn^2}[\Sigma_{S}(\t)]\right\}, \\
\Pi_{S_2}(\t)
&=&-\lambda_{B}^2\left\{{1\over(\t-\mn^2)^3}[-8\mn\Sigma_{1}(\t)]
+{1\over(\t-\mn^2)^2}[\mn\Sigma_{2}(\t)+\mn^3\Sigma_{3}(\t)+2\Sigma_{V}(\t)]
\right.\cr
& &\left.\qquad
+{1\over\t-\mn^2}[\mn\Sigma_{3}(\t)]\right\}, \\
\Pi_{V_1}(\t)
&=&-\lambda_{B}^2\left\{
    {1\over(\t-\mn^2)^3}[4\mn^2\Sigma_{1}(\t)]
+{1\over(\t-\mn^2)^2}[\Sigma_{1}(\t)+2\mn\Sigma_{S}(\t)]\right\},\\
\Pi_{V_2}(\t)
&=&-\lambda_{B}^2\left\{
    {1\over(\t-\mn^2)^2}[4\Sigma_{1}(\t)-\mn^2\Sigma_{2}(\t)]
+{1\over\t-\mn^2}[-\Sigma_{2}(\t)-\frac{\Sigma_{V}(\t)}{\mn}]\right\},\\
\Pi_{V_3}(\t)
&=&-\lambda_{B}^2\left\{
    {1\over(\t-\mn^2)^3}[-8\Sigma_{1}(\t)]
+{1\over(\t-\mn^2)^2}[2\Sigma_{2}(\t)
     +\mn^2\Sigma_{3}(\t)+\frac{2\Sigma_{V}(\t)}{\mn}]\right.\cr
& &\qquad\left.+{1\over\t-\mn^2}[\Sigma_{3}(\t)]\right\}.
\eey
$\Sigma_{V}$ and $\Sigma_{S}$ on the mass-shell are related
with the polarizabilities $\bar\alpha_B$ and $\bar\beta_B$:
\ben
\Sigma_{V}(\mn^2)=(\bar\alpha_B+\bar\beta_B)/2, \qquad
\Sigma_{S}(\mn^2)=-\bar\beta_B/4, \label{se-pol}
\een
while $\Sigma_{1}$, $\Sigma_{2}$ and $\Sigma_{3}$
with other quantities [\ref{QSR}].
Defining the following combination: 
\ben
\Pi^{(\bar\alpha_B)}(p^2)=
-2\Pi_{S_1}(p^2)-p^2\Pi_{S_2}(p^2)+2\mn\Pi_{V_1}(p^2)
+\mn\Pi_{V_2}(p^2)+p^2\mn\Pi_{V_3}(p^2),
\een
we find the imaginary part of $\Pi^{(\bar\alpha_B)}$ to be
related solely with $\bar\alpha_B$:
\ben
{1\over\pi}{\rm Im}\;\Pi^{(\bar\alpha_B)}(\t)
 =-\lambda_{B}^2\delta(\t-\mn^2){\bar\alpha_B\over2}. \label{imagin}
\een
For the magnetic polarizabilities ${\bar\beta}_{B}$, it seems difficult to 
derive relations free from the constants
which are not known well.
Numerical estimation of ${\bar\beta}_B$ is therefore left for a future work.

We write the interpolating fields for the 
$n$, $\Xi^0$, $\Sigma^0$, and $\Lambda$, respectively, as follows [\ref{yazaki}]:
\bey
\eta_{n}(x)&=&\epsilon_{abc}[d^{aT}(x)C\gamma_{\mu}d^{b}(x)]
                            \gamma_{5}\gamma^{\mu}u^{c}(x), \label{ncurrent}\\
\eta_{\Xi^0}(x)&=&-\epsilon_{abc}\left[s^{aT}(x)C\gamma_{\mu}s^{b}(x)\right]
                            \gamma_{5}\gamma^{\mu}u^{c}(x), \label{xi0current}\\
\eta_{\Sigma^0}(x)&=&\sqrt{2}\epsilon_{abc}
                   \left\{\left[u^{aT}(x)C\gamma_{\mu}s^{b}(x)\right]
                  \gamma_{5}\gamma^{\mu}d^{c}(x) 
                         +\left[d^{aT}(x)C\gamma_{\mu}s^{b}(x)\right]
                  \gamma_{5}\gamma^{\mu}u^{c}(x)\right\}, \label{sigma0current}\\
\eta_{\Lambda}(x)&=&\sqrt{\frac{2}{3}}\epsilon_{abc}
                   \left\{\left[u^{aT}(x)C\gamma_{\mu}s^{b}(x)\right]
                  \gamma_{5}\gamma^{\mu}d^{c}(x) 
                  -\left[d^{aT}(x)C\gamma_{\mu}s^{b}(x)\right]
                  \gamma_{5}\gamma^{\mu}u^{c}(x)\right\}, \label{lambdacurrent}
\eey
where $C$ denotes the charge conjugation and $a$, $b$ and $c$ color indices.
Following Ref.[\ref{QSR}] the calculations of the Wilson coefficients 
are straightforward.

Let us list the vacuum expectation values of interest and then
classify them according to their dimensions.
We are interested in the quadratic terms in $\F$. Hence, the
vacuum expectation value (VEV) of the lowest-dimensional operator
${\bar q}q$
is the second order term of the quark condensate in the external
field, 
when we expand it in powers of $F_{\mu\nu}$
\footnote{In a recent work [\ref{Smilga}], it is argued that $\langle{\bar q}q
\rangle_{F}$ cannot be expanded in powers of $F_{\mu\nu}$ for massless
quarks. We find, however, that $\langle{\bar q}q\rangle_{F}$
can be expanded in the NJL model for massless quarks [\ref{NJL}]. 
In this paper we assume that the expansion is possible.}.
Four-dimensional operators are
$F_{\mu\nu}F_{\rho\sigma}$ itself, $i{\bar q}{\cal S}\gamma_{\mu}D_{\nu}q$
and $G_{\mu\nu}^{a}G_{\rho\sigma}^{a}$,
where the covariant derivative is given by
$D_{\mu}=\partial_{\mu}+ig(\lambda^{a}/2)A_{\mu}^{a}$,
and ${\cal S}$ is a symbol which makes the operators symmetric 
and traceless with respect to the Lorentz indices.
The contribution of the VEV of the last two operators are so
small that they can be safely neglected [\ref{QSR}].
The following three operators: 
${\bar q}\sigma_{\mu\nu}q\cdot F_{\rho\sigma}$, 
$g{\bar q}\sigma_{\mu\nu}(\lambda^{a}/2)G_{\rho\sigma}^{a}q$,
${\bar q}{\cal S}D_{\mu}D_{\nu}q$
have dimension 5.
The contribution of the second operator turns out to vanish, as is for
the sum rule for the masses of octet baryons.
The Wilson coefficient of the last operator is constant 
and vanishes after a Borel transform.
There are three six-dimensional operators; that is,
$g{\bar q}\gamma_{\mu}D_{\nu}(\lambda^{a}/2)G_{\rho\sigma}^{a}$,
${\bar q}q{\bar q}q$ and 
${\bar q}\sigma_{\mu\nu}q{\bar
q}\sigma_{\rho\sigma}q$.
The VEV of the first operator cannot be induced
by the second-order interaction of the vacuum with an electromagnetic field,
since 
${\bar q}\gamma_{\mu}D_{\nu}(\lambda^{a}/2)G_{\rho\sigma}^{a}q$
has odd C-parity.
To estimate the VEV's of the last two operators we have applied
the factorization hypothesis, which is firmly based in the case of
the four quark operators:
$\langle{\bar q}q{\bar q}q\rangle_{F}
=2\langle{\bar q}q\rangle_{F}\cdot\langle{\bar q}q\rangle_{0}$,
$\langle{\bar q}\sigma_{\mu\nu}q{\bar
q}\sigma_{\rho\sigma}q\rangle_{F}
=\langle{\bar q}\sigma_{\mu\nu}q\rangle_{F}\cdot\langle{\bar
q}\sigma_{\rho\sigma}q\rangle_{F}$, 
where $\langle\cdots\rangle_F$ denotes the VEV in the electromagnetic field
and $\langle\cdots\rangle_0$ the VEV when the external field is absent.   

For the calculation up to six-dimensional terms, 
we are then left with the two external-field-induced VEV's:
$\langle{\bar q}\sigma_{\mu\nu}q\rangle_{F}$,
$\langle{\bar q}q\rangle_{F}$.
It is assumed in Ref.[\ref{I&S}] that the first of them can be written, in
general,
\ben
  \langle{\bar q}\sigma_{\mu\nu}q\rangle_{F}
      =\sqrt{4\pi\alpha} F_{\mu\nu} \chi e_{q} \langle{\bar q}q\rangle_{0},
\een
where $\alpha$ is the fine structure constant and 
$\chi$ the so-called magnetic susceptibility of quark condensate.
We extend this hypothesis to the case of the second order in $F_{\mu\nu}$
and define a ``susceptibility'' $\phi$ as follows:
\ben
\qqf=(1+4\pi\alpha F_{\mu\nu}F^{\mu\nu} \phi e_{q}^2 )
                     \qqv.\label{phi} 
\een

In order to obtain the desired sum rule, 
we substitute Eq.(\ref{imagin}) into Eq.(\ref{BSR}).
The contributions to the imaginary parts from higher mass states are
approximated using the logarithmic terms in the OPE, 
starting at a threshold $S_0$. 
We calculate the left hand side of Eq.(\ref{BSR}) by an OPE up to
dimension 6.
For hyperons, we further introduce the effect of the
finite mass of the strange quark, $m_s$.
In the sum rule for $\bar\alpha_{\Xi^0}$ 
the leading terms of the mass correction appear in 
the dimension 6 terms.
In the sum rules for $\bar\alpha_\Lambda$ and $\bar\alpha_{\Sigma^0}$ 
we consider only the leading correction, which appears
in the dimension 4 terms, namely $m_s\langle{\bar q}q\rangle$.
One finally obtain the following results:
\bey
& & \frac{\alpha}{4\pi^3}\Big\{
     \Big[8\pi^2 \phi e_{u}^2 \uuv s^2E_1(S_0,s)
          +\frac{4}{3}\pi^2\{\chi(6 e_{u}e_d+ e_{u}^2)\}
          \uuv sE_0(S_0,s)\Big] \nnbr \\
& &\qquad   -M_\lambda
     \Big[ 2e_{u}e_{d}sE_0(S_0,s)
          -\frac{32}{3}\pi^4\{4 \phi e_{d}^2 - \chi^2 e_{d}^2\}(\ddv)^2
       \Big]\Big\} \nnbr \\
&=&-\frac{1}{2}\lambda_n^2\bar\alpha_n e^{-M_n^2/s},  \label{BSRalphan}
\eey
\bey
& &\frac{\alpha}{4\pi^3}\Big\{
    \Big[8\pi^2 \phi e_{u}^2\uuv s^2E_1(S_0,s)
     +\frac{4}{3}\pi^2\{\chi(6e_{u}e_s+e_{u}^2)\}
        \uuv sE_0(S_0,s) \Big] \nnbr \\
& &\qquad  -M_\Xi
       \Big[ 2 e_{u}e_{s}sE_0(S_0,s)
        -\frac{32}{3}\pi^4\{4\phi e_{s}^2 - \chi^2 e_{s}^2\}(\ssv)^2
       \Big]\Big\} \nnbr \\
&=&-\frac{1}{2}\lambda_\Xi^2\bar\alpha_{\Xi^0} e^{-M_\Xi^2/s}
        , \label{BSRalphaxi0}
\eey
\bey
& &\frac{\alpha}{4\pi^3}\Big\{
    \Big[8\pi^2 \phi e_s^2\ssv s^2E_1(S_0,s)  
         +\frac{4}{3}\pi^2\chi(3e_u e_s+3e_d e_s+e_s^2)
           \ssv sE_0(S_0,s)\Big] \nnbr \\
& &\qquad -M_\Sigma
  \Big[ \{(e_u+e_d)e_s-8\pi^2 \phi m_s e_s^2 \ssv\}sE_0(S_0,s) 
 \nnbr \\
& &\qquad
  -\frac{32}{3}\pi^4\{2\phi(e_u^2+e_d^2) - \chi^2 e_u e_d\}\uuv\ddv
       \Big]\Big\} \nnbr \\
&=&-\frac{1}{2}\lambda_\Sigma^2\bar\alpha_{\Sigma^0} e^{-M_\Sigma^2/s}
        , \label{BSRalphasigma0}
\eey
\bey
& &\frac{\alpha}{12\pi^3}\Big\{
\Big[8\pi^2\phi\{2e_u^2  \uuv+2e_d^2\ddv-e_s^2\ssv\} s^2E_1(S_0,s) 
 \nnbr \\
& &\qquad\quad
     +\frac{4}{3}\pi^2\chi\{2(e_u+3e_d+3e_s)e_u\uuv+2(e_d+3e_u+3e_s)e_d\ddv
 \nnbr \\
& &\qquad\qquad\qquad -(e_s+3e_u+3e_d)e_s\ssv\} sE_0(S_0,s)
 \Big] \nnbr \\
& &\quad -M_\Lambda
\Big[\{e_s e_u+e_s e_d+4e_u e_d
   +8\pi^2 \phi m_s(2e_u^2\uuv+2e_d^2\ddv-3e_s^2\ssv)\}sE_0(S_0,s)
\nnbr \\
& &\qquad\quad
-\frac{64}{3}\pi^4
 \phi\{2(e_u^2+e_s^2)\uuv\ssv+2(e_d^2+e_s^2)\ddv\ssv-(e_d^2+e_u^2)\ddv\uuv\}
\nnbr \\
& &\qquad\quad
+\frac{32}{3}\pi^4
  \chi^2\{2e_u e_s\uuv\ssv+2e_d e_s\ddv\ssv-e_u e_d\uuv\ddv\}
       \Big]\Big\} \nnbr \\
&=&-\frac{1}{2}\lambda_\Lambda^2\bar\alpha_\Lambda
e^{-M_\Lambda^2/s},
\label{BSRalphalambda}
\eey
where $E_{n}(S_{0},s)=1-{\rm exp}(-S_0/s)\sum_{k=0}^{n}(S_0/s)^k/k!$.
It is well known that from $\Pi^{(0)}(p)$ in Eq.(\ref{tenkai}) one can 
get the Borel sum rules
for $\lambda_{B}^2$ listed below.
\ben
s^3E_{2}(S_{0},s)
+\pi^2\langle\frac{\alpha_s}{\pi}G^2\rangle_0
s E_{0}(S_{0},s)+ \frac{64}{3}\pi^4(\ddv)^2
=2(2\pi)^4\lambda_{N}^2{\rm e}^{-M_N^2/s}, \label{res-n} 
\een
\ben 
s^3
E_{2}(S_{0},s)+\pi^2\langle\frac{\alpha_s}{\pi}G^2\rangle_0
s E_{0}(S_{0},s)+ \frac{64}{3}\pi^4(\ddv)^2
=2(2\pi)^4\lambda_{\Xi}^2{\rm e}^{-M_{\Xi}^2/s}, \label{res-xi} 
\een
\ben
s^3 E_{2}(S_{0},s)
+8\pi^2 m_s \ssv s E_{0}(S_{0},s)
+\pi^2\langle\frac{\alpha_s}{\pi}G^2\rangle_0 
s E_{0}(S_{0},s)
+\frac{64}{3}\pi^4(\uuv)^2
=2(2\pi)^4\lambda_{\Sigma}^2{\rm e}^{-M_{\Sigma}^2/s} ,
\label{res-sigma} 
\een
\bey
&&s^3 E_{2}(S_{0},s)
-\frac{8}{3}\pi^2 m_s[2(\uuv+\ddv)-3\ssv]sE_{0}(S_{0},s)
+\pi^2\langle\frac{\alpha_s}{\pi}
  G^2\rangle_0 s E_{0}(S_{0},s) \nnbr \\
&&
+\frac{64}{9}\pi^4[2(\uuv+\ddv)\ssv-\uuv\ddv]
=2(2\pi)^4\lambda_{\Lambda}^2{\rm e}^{-M_{\Lambda}^2/s},
\label{res-lambda}
\eey
where  
$\langle (\alpha_s/\pi)G^2\rangle_0$ denotes the gluon condensate.

Here, let us compare the sum rules for $\bar\alpha_B$, 
Eqs.(\ref{BSRalphan})$\sim$(\ref{BSRalphalambda}).
In the $n$ and $\Xi^0$ the left-hand sides are similar to one another
because of the same isospin-$1/2$ structure of the interpolating fields in
Eqs.(\ref{ncurrent}) and (\ref{xi0current}), while the interpolating
fields of Eqs.(\ref{sigma0current}) and (\ref{lambdacurrent}) have
isospin-$1$ and $0$, respectively and the above similarity is not seen.
In the limit of flavor SU(3) symmetry, the $n$ and $\Xi^0$ exactly
coincide with each other, because of $e_s=e_d$.
Furthermore, up to dimension 5 we find the generic results:
\ben
\bar\alpha_n=\bar\alpha_{\Xi^0}
=\frac{4}{3}\bar\alpha_{\Lambda}=4\bar\alpha_{\Sigma^0}.
\een
It is interesting to know 
that not only the leading terms but also higher dimensional terms 
satisfy the above relation.

Let us now estimate $\bar\alpha_B$ numerically.
We eliminate $\lambda_{B}^2$ 
in Eqs.(\ref{BSRalphan})$\sim$(\ref{BSRalphalambda}) 
by using Eqs.(\ref{res-n})$\sim$(\ref{res-lambda}).
For the susceptibilities we use the values evaluated in
Ref.[\ref{QSR}]: 
\bey
 \chi=-2.58\,\,{\rm GeV}^{-2},\quad
\phi=1.66\,\,{\rm GeV}^{-4}.
\eey
We note, however, that there is an uncertainty
due to the variation of the susceptibilities with the change of $s$:
In the reliable interval of $s$, $\chi$ varies from
$-2.76\;{\rm GeV}^{-2}$ to $-2.44\;{\rm GeV}^{-2}$
,$\phi$ from $1.49\;{\rm GeV}^{-4}$ to $1.91\;{\rm GeV}^{-4}$
. The vacuum condensates of the operators and the mass of the strange
quark are taken from Ref.[\ref{yazaki}], 
and are
\begin{eqnarray*}
\uuv=\ddv=(-225\,{\rm MeV})^3,\,\ssv=0.8\uuv,\,
\langle \frac{\alpha_s}{\pi}G^2\rangle_0=(360\,{\rm MeV})^4,
\,m_s=150\,{\rm MeV}.
\end{eqnarray*}
In Fig.1 we show the squared Borel mass $s$ dependence of $\bar\alpha_B$.
We see that, for $\bar\alpha_n$, $\bar\alpha_{\Xi^{0}}$ and
$\bar\alpha_{\Lambda}$, 
the plateau develops for the 
suitable values of the continuum threshold parameter $S_0$.
This implies that the sum rules for $\bar\alpha_n$,
$\bar\alpha_{\Xi^{0}}$ and $\bar\alpha_\Lambda$ work very well
and are reliable.
Indeed, we see that dominant contributions in the OPE come from the leading terms
and the higher dimensional terms are suppressed.
The convergence of the OPE hence appears to be good.
Taking the maximum value of the curves, we predict that 
\bey
\bar\alpha_n&=&14.0\times 10^{-4}{\rm fm}^3, \\
\bar\alpha_{\Xi^{0}}&=&10.8\times 10^{-4}{\rm fm}^3,    \\
\bar\alpha_{\Lambda}&=&6.2\times 10^{-4}{\rm fm}^3.
\eey
These results have uncertainties arising from the errors in the
susceptibilities. With the variation of the susceptibilities,
$\bar\alpha_n$ varies from $12.1$ to $16.8$, 
$\bar\alpha_{\Xi^{0}}$ $9.1$ to $13.5$, 
and $\bar\alpha_{\Lambda}$ $5.0$ to $8.0$. 

On the other hand, the Borel curve of $\bar\alpha_{\Sigma^{0}}$
are not stabilized for any choice of $S_0$ due to apparently slow
convergence of the OPE. Indeed, the dimension 6 term is significantly
large and contributes to $\bar\alpha_{\Sigma^{0}}$ negatively.
In view of this, no definite conclusion for $\bar\alpha_{\Sigma^{0}}$
can be drawn. However, we can say that $\bar\alpha_{\Sigma^0}$
might take a small value.

It is interesting to compare our results with those obtained
in the heavy baryon chiral perturbation theory (HBChPT) [\ref{HBCHPT}].
The QCD sum rules give smaller values for the electric polarizabilities of 
hyperons as compared
with that of neutron. This is in agreement with the HBChPT prediction.
In the HBChPT, as already mentioned, the splitting is due to the contribution
of kaon cloud.
It can be seen that the ordering of magnitudes for strange baryons
is reversal in the
two calculations; namely, in the HBChPT we find $\bar\alpha_{\Sigma^0}>
\bar\alpha_\Lambda>\bar\alpha_{\Xi^0}$, while $\bar\alpha_{\Xi^0}>
\bar\alpha_\Lambda>\bar\alpha_{\Sigma^0}$ in the QCD sum rules. 
It is known that the effect of the decouplet states is large, so that
it is highly desired to obtain the result of the HBChPT with including 
the decouplet states [\ref{Hemmert}]. 

For charged baryons, there newly appear in the phenomenological side 
a triple pole term which is proportional to squared charge of the
baryon and involves large factor. 
Within the calculation up to dimension 6 in the OPE, 
this term makes the Borel stability worse.
We therefore need to evaluate the higher dimensional terms.
This task is presently under way. 

In summary, we have presented a calculation of the electric polarizabilities of
neutral baryons in the framework of the QCD sum rule. The operators 
up to dimension 6 were retained in the operator product expansion.
The sum rule predicts that the ordering of the polarizabilities is
$\bar\alpha_n>\bar\alpha_{\Xi^0}>\bar\alpha_{\Lambda}>\bar\alpha_{\Sigma^0}$,
and implies that $\bar\alpha_{\Sigma^0}$ has a possibility
to be very small. In particular, we find that in the flavor
SU(3) symmetric limit, 
$\bar\alpha_n=\bar\alpha_{\Xi^0}
=4\bar\alpha_{\Lambda}/3=4\bar\alpha_{\Sigma^0}$, 
up to dimension 5 terms in the OPE.
The calculations performed in the HBChPT have led still
different predictions. However, the future experimental data from CERN 
and Fermilab could be of help to discriminate the different approaches.

One of the authors (Y.K.) would like to thank O. Morimatsu,
M. Oka and D. Jido for helpful discussions.
Y.K. thanks also the members of the theory group of Institute of
Particle and Nuclear Studies for their hospitality
during his stay there.


\newpage
\baselineskip 24pt
\begin{center}
{\bf References}
\end{center}
\def\labelenumi{[\theenumi]}
\def\Ref#1{[\ref{#1}]}
\def\npb#1#2#3{{Nucl. Phys.\,}{\bf B{#1}}\,(#3), #2}
\def\npa#1#2#3{{Nucl. Phys.\,}{\bf A{#1}}\,(#3),#2}
\def\np#1#2#3{{Nucl. Phys.\,}{\bf{#1}}\,(#3),#2}
\def\plb#1#2#3{{Phys. Lett.\,}{\bf B{#1}}\,(#3),#2}
\def\prl#1#2#3{{Phys. Rev. Lett.\,}{\bf{#1}}\,(#3),#2}
\def\prd#1#2#3{{Phys. Rev.\,}{\bf D{#1}}\,(#3),#2}
\def\prc#1#2#3{{Phys. Rev.\,}{\bf C{#1}}\,(#3),#2}
\def\pr#1#2#3{{Phys. Rev.\,}{\bf{#1}}\,(#3),#2}
\def\ap#1#2#3{{Ann. Phys.\,}{\bf{#1}}\,(#3),#2}
\def\prep#1#2#3{{Phys. Reports\,}{\bf{#1}}\,(#3),#2}
\def\rmp#1#2#3{{Rev. Mod. Phys.\,}{\bf{#1}}\,(#3),#2}
\def\cmp#1#2#3{{Comm. Math. Phys.\,}{\bf{#1}}\,(#3),#2}
\def\ptp#1#2#3{{Prog. Theor. Phys.\,}{\bf{#1}}\,(#3),#2}
\def\ib#1#2#3{{\it ibid.\,}{\bf{#1}}\,(#3),#2}
\def\zsc#1#2#3{{Z. Phys. \,}{\bf C{#1}}\,(#3),#2}
\def\zsa#1#2#3{{Z. Phys. \,}{\bf A{#1}}\,(#3),#2}
\def\intj#1#2#3{{Int. J. Mod. Phys.\,}{\bf A{#1}}\,(#3),#2}
\def\sjnp#1#2#3{{Sov. J. Nucl. Phys.\,}{\bf #1}\,(#3),#2}
\def\pan#1#2#3{{Phys. Atom. Nucl.\,}{\bf #1}\,(#3),#2}

\def\etal{{\it et al.}}
\begin{enumerate}

\divide\baselineskip by 4
\multiply\baselineskip by 3

\item\label{Lvov}
A. I. L'vov, \intj{8}{5267}{1993}.
\item \label{exp:Fed}
F. J. Federspiel et al., \prl{67}{1511}{1991}.
\item \label{exp:Zie}
A. Zieger et al., \plb{278}{34}{1992}.
\item \label{exp:Hal}
E. Hallin et al., \prc{48}{1497}{1993}.
\item \label{exp:Sch}
J. Schmiedmeyer et al., \prl{66}{1015}{1991}.
\item \label{Felmilab}
M. A. Moinester, Proc. Workshop on Chiral Dynamics, July 1994,
eds. A. Bernstein and B. Holstein, Los Alamos archive hep-ph/9409463.
\item \label{NRQM}
H. J. Lipkin and M. A. Moinester, \plb{287}{179}{1992}.
\item \label{HBCHPT}
V. Bernard, N. Kaiser, J. Kambor and Ulf-G. Meissner,
\prd{46}{R2756}{1992}; \plb{319}{269}{1993}.
\item \label{BS-skyrme}
C. Gobbi, C. L. Schat and N. N. Scoccola, \npa{598}{318}{1996}
\item \label{SVZ1} M. A. Shifman, A. I. Vainshtein and~V. I. Zakharov, \npb{147}{385}{1979}.
\item \label{SVZ2} M. A. Shifman, A. I. Vainshtein and~V. I. Zakharov,
\npb{147}{448}{1979}.
\item \label{QSR}
T. Nishikawa, S. Saito and Y. Kondo, \npa{615}{417}{1997}.
\item \label{I&S}
B. L. Ioffe and A. V. Smilga, \npb{232}{109}{1984}.
\item \label{yazaki}
L. J. Reinders, H. R. Rubinstein, and S. Yazaki,
\prep{127}{1}{1985}.
\item \label{Smilga}
I. A. Shushpanov and A. V. Smilga, \plb{402}{351}{1997}.
\item \label{NJL}
S. P. Klevansky, \rmp{64}{649}{1992} and
references therein.
\item \label{Hemmert}
T. R. Hemmert, B. R. Holstein and J. Kambor, \prd{55}{5598}{1997}.
\end{enumerate}
\newpage
\centerline{\bf Figure Caption}
\noindent
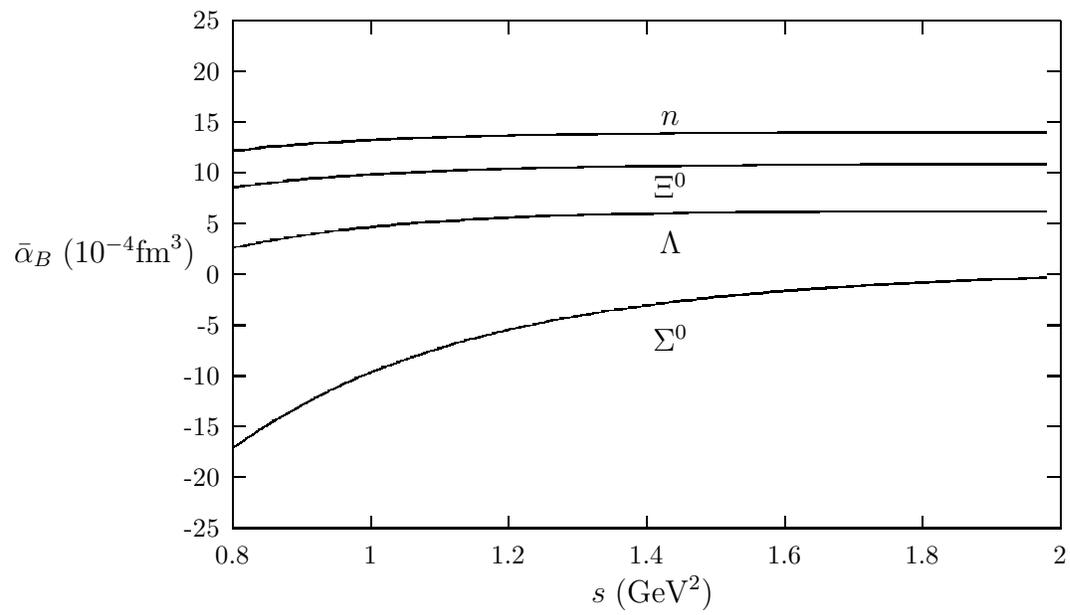
{\bf Fig.1}\\
The squared Borel mass $s$ dependence of ${\bar\alpha}_{B}$. 
The values of the continuum threshold parameter $S_0$
are taken to be $3.6\;{\rm GeV}^{2}$ for the $n$, 
$4.2\;{\rm GeV}^{2}$ for the $\Xi^0$,
$6.7\;{\rm GeV}^{2}$ for the $\Lambda$ and $\infty\;{\rm GeV}^{2}$
for the $\Sigma^0$.

\clearpage
\pagestyle{empty}
\begin{figure}
\begin{center}
\input{neutBalpha-s-nomixedcon}
\\
\vspace{1cm}
{\bf Fig.1}
\end{center}
\end{figure}

\end{document}